\documentclass[twocolumn,showpacs,preprintnumbers,amsmath,amssymb]{revtex4}

\usepackage{graphicx}
\usepackage{dcolumn}
\usepackage{bm}

\begin{document}


\title{Photon detection by current-carrying superconducting film: A time-dependent Ginzburg-Landau
approach}

\author{A. N. Zotova}
\author{D. Y. Vodolazov}
\email{vodolazov@ipm.sci-nnov.ru} \affiliation{Institute for
Physics of Microstructures, Russian Academy of Sciences, 603950,
Nizhny Novgorod, GSP-105, Russia}

\date{\today}


\begin{abstract}
We study dynamics of the order parameter in a superconducting film
with transport current after absorption of a single photon. The
system from the time-dependent Ginszburg-Landau equation,
Poisson's equation for an electrical potential and the heat
diffusion equation was solved numerically. For each photon energy
in the absence of fluctuations, there exists a corresponding
threshold current below which the superconducting state is stable
and no voltage appears between the ends of the film. At larger
currents the superconducting state collapses starting from the
appearance of a vortex-antivortex pair in the center of the region
with suppressed superconducting order parameter which has been
created by the absorbed photon. Lorentz force causes motion of
these vortices that heats the film locally and gives rise to a
normal domain. When biased with the fixed current, the film
latches in the normal state. In the regime when current via
superconductor may change, which is more relevant for experiments,
the normal domain exists only for a short time resulting in the
voltage pulse with the duration controlled by the kinetic
inductance of the superconducting film.
\end{abstract}

\maketitle

\section{Introduction}

Despite the large number of both experimental (see for example
\cite{Gol'tsman,Verevkin,Tarkhov,Divochiy,Yang1,Shibata,Hofherr}
and references therein) and theoretical
\cite{Semenov1,Semenov2,Yang2,Jukna} works on the superconductive
single-photon detectors (SSPD) still there are some questions
about mechanism of photon detection by superconductive film
carrying the transport current. The original understanding of the
detection mechanism is following - after absorption of the single
photon the hot spot is formed in the superconducting film. It
locally destroys superconductivity and leads to concentration of
the current density outside the hot spot due to decrease of the
effective width of the film \cite{Semenov1}. If transport current
is close to the depairing current then the current density outside
the hot spot area may exceed the depairing current density and the
superconducting state becomes unstable, leading to the voltage
response.

The quantitative analysis of initial stage of the hot spot
formation in the existing theoretical models is based on the
solution of the diffusion equation for nonequilibrium
quasiparticles \cite{Semenov1,Semenov2,Jukna}. It was postulated
that when the number of the nonequilibrium quasiparticles in the
hot spot exceeds some critical value this region could be
considered as a normal one and the superconducting current is
forced to flow around it. In the refined model \cite{Semenov2} it
was supposed that even partial suppression of the superconducting
order parameter in the hot spot leads to current enhancement
outside that region and to instability of the superconducting
state and formation of the normal domain. The further evolution of
the normal domain is usually studied by using the heat diffusion
equation for effective temperature of quasiparticles coupled with
equation describing the embedding circuit
\cite{Yang2,Semenov5,Kerman}.

In the majority of previous models it was implicitly assumed that
the magnitude of the superconducting order parameter $|\Delta|$
changes instantly in time if local temperature $T(\vec{r},t)>T_c$
or local superconducting current density $j(\vec{r},t)$ exceeds
depairing current density $j_{dep}$. But it is well know that
$|\Delta|$ has finite relaxation time $\tau_{|\Delta|}$ and in
some cases $\tau_{|\Delta|}$ could be comparable with
electron-phonon inelastic relaxation time $\tau_{e-ph}$
\cite{Tinkham}. Because energy relaxation of nonequilibrium
quasiparticles in the hot spot occurs on the same time scale (or
even much shorter due to diffusion of the quasiparticles) it is
clear that finite $\tau_{|\Delta|} \neq 0$ should influence the
photon detection process \cite{Zhang}.

Another interesting and unresolved question is what kind of
instability of the superconducting state occurs due to appearance
of the hot spot region. Is it gradual suppression of the order
parameter outside the hot spot due to current concentration
\cite{Semenov1,Zhang} or nucleation of the vortex-antivortex pair
inside the hot spot \cite{Kadin}?

In our work we use the simplest approach where effect of finite
$\tau_{|\Delta|}$ is taken into account and stability of the
superconducting state is analyzed self-consistently. The dynamics
of the superconducting order parameter is studied on the basis of
the time-dependent Ginzburg-Landau equation. This equation is
coupled with the heat diffusion equation for the effective
temperature of the quasiparticles and Poisson's equation for the
electrical potential. We consider the current bias regime as well
as the regime when current via superconductor may change due to
presence of the shunt resistance and take into account the finite
kinetic inductance of the film.

Within this model we show that incoming photon creates the finite
size region with partially suppressed order parameter. We find
that even for {\it infinite} superconducting film such a state
becomes unstable {\it without any fluctuations} with respect to
appearance of the vortex-antivortex pair (or single vortex if
photon is absorbed on the edge of the film) at threshold current
{\it less than depairing current}. Motion of the vortex and
antivortex in opposite directions under the Lorentz force heats
the sample (if the threshold current is not too small). As a
result, the normal domain appears which either expands over the
whole film (current bias regime) or shrinks and disappears (when
current via superconductor may change) resulting in the voltage
pulse. Our result supports the hypothesis of Ref. \cite{Kadin}
that the single photon can nucleate the vortex-antivortex pair in
current-carrying superconductor and their motion provides the
voltage pulse which could be detected. Our model also confirms the
experimentally observed smeared red boundary in the single-photon
detection.

The paper is organized as follows. In section II we present the
theoretical model. The results of the numerical calculations and
simple analytical estimations are given in Section III. In section
IV we discuss the relation of our results with an experiment and
in section V we present our conclusions.

\section{Model}

In our work we do not study the initial part of the detection
process when the single photon is absorbed by the electron. We use
approach of the effective temperature \cite{Giazotto}, which is
valid when the thermalization time (which is proportional to the
electron-electron inelastic relaxation time $\tau_{e-e}$) is
shorter than the inelastic relaxation time due to electron-phonon
interactions $\tau_{e-ph}$. We assume that during initial time
interval $\sim \tau_{e-e}$ after absorption of the photon the
electron-electron interactions creates hot spot with radius
$R_{init}\sim L_{e-e}=(D\tau_{e-e})^{1/2}$ (D is a diffusion
constant) and with local temperature $T_0+\Delta T$ ($T_0$ is a
bath temperature) where $\Delta T$ is determined from the energy
conservation
\begin{equation}
2\pi \hbar c/\lambda=\Delta T \pi R_{init}^2 d C_v
\end{equation}

Here $\lambda$ is a wavelength of the electromagnetic radiation,
$\hbar$ is a Planck constant, c is a speed of light, d is a
thickness of the film and $C_v$ is a heat capacity of the
quasiparticles (for simplicity we take $C_v$ as in the normal
state at $T=T_c$).

Time and space evolution of the temperature in the superconducting
film we find from the heat diffusion equation
\begin{equation}
\frac{\partial T}{\partial t}=D\left(\frac{\partial^2 T}{\partial
x^2}+\frac{\partial^2 T}{\partial y^2}\right)+\frac{\rho_n
j_n^2}{C_v}-\frac{T-T_0}{\tau_{e-ph}}
\end{equation}
where $\rho_n$ is a normal state resistivity, $j_n=-\nabla
\varphi/\rho_n$ is a normal current density and $\varphi$ is a
electrostatic potential. Here we assume that the phonons are in
equilibrium with the bath and energy relaxation occurs due to
interaction with phonons. Our calculations show, that initial
destruction of superconductivity occurs on timescale shorter than
$\tau_{e-ph}$ and therefore at initial stage of dynamical response
of $|\Delta|$ one may neglect the heating of phonons (in our model
we neglect possibility of the phonon heating during initial
$t\lesssim \tau_{e-e}$ stage of hot spot formation).

To study the dynamics of the order parameter
$\Delta=|\Delta|e^{i\phi}$ we use the time-dependent
Ginzburg-Landau equation
\begin{eqnarray}
\nonumber \frac{\pi \hbar}{8k_BT_c}\left(\frac{\partial }{\partial
t}-\frac{i2e \varphi}{\hbar} \right)\Delta=\xi_{GL}(0)^2
\left(\frac{\partial^2\Delta}{\partial
x^2}+\frac{\partial^2\Delta}{\partial y^2}\right)
\\
+\left(1-\frac{T}{T_c}-\frac{|\Delta|^2}{\Delta_{GL}(0)^2}\right)
\Delta,
\end{eqnarray}
where $\xi_{GL}(0)=(\pi\hbar D/8k_BT_c)^{1/2}$ and
$\Delta_{GL}(0)=4k_BT_c u^{1/2}/\pi$ ($u \simeq 5.79$ - see
\cite{Kramer}) are the zero temperature Ginzburg-Landau coherence
length and the order parameter respectively. Characteristic time
relaxation of the order parameter described by Eq. (3) is
$\tau_{|\Delta|}=\pi\hbar/8k_B(T_c-T)$. Although Eq. (3) is
quantitatively valid only near critical temperature of the
superconductor (at $T \gtrsim 0.9 T_c$ when $\tau_{e-e}\ll
\tau_{e-ph}$ and $\tau_{e-e}\ll \tau_{|\Delta|}$) we use it to
model the dynamics of the superconducting condensate at lower
temperatures to find some qualitative results.

We should complete Eqs. (2-3) by equation for the electric
potential $\varphi$ which comes from the conservation of the full
current $div (j_s+j_n)=0$
\begin{equation}
\Delta \varphi =\rho_n div (j_s)
\end{equation}
here $j_s={\rm Imag}(\Delta^*\nabla \Delta)/(4ek_BT_c\rho_n)$.
\begin{figure}[hbtp]
\includegraphics[width=0.4\textwidth]{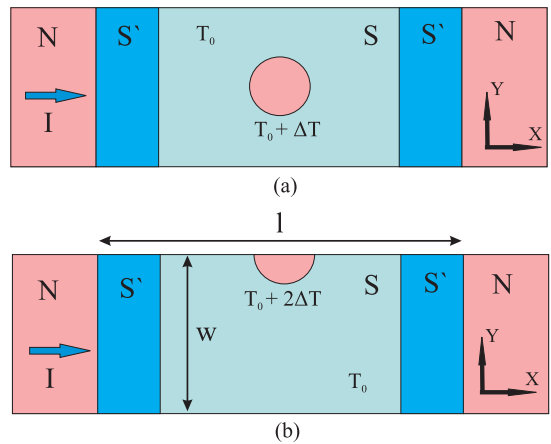}
\caption{The model geometry: the superconductive film is placed
between two normal bulk contacts; (a) - the photon is absorbed in
the center of the film, (b) - the photon is absorbed on the edge
of the film.}
\end{figure}

To model the response of the superconducting film after absorption
of the single photon we consider the model geometry which is
present in Fig. 1. We need the normal contacts ($\Delta=0$,
$\partial \varphi/\partial x=- \rho_n I/(wd)$ at $x=\pm l/2$) to
inject the current to the superconducting film in our numerical
calculations and which are kept at the bath temperature $T_0$
($T|_{x=\pm l/2}=T_0$). The current and heat do not flow through
the lateral edges of the film ($\partial T/\partial y=0$,
$\partial \varphi/\partial y =0$, $\partial \Delta/\partial y =0$
at $y=\pm w/2$). To neglect the influence of the N-S boundaries
(for example the motion of the NS boundary) on the dynamical
processes in the superconducting film we artificially enhance the
superconducting order parameter in the regions marked by dark blue
color in Fig. 1 by introducing locally higher $T_c$ (the width of
these regions is larger than the penetration depth of the electric
field from the normal contact and is equal to 5$\xi_{GL}(0)$).

\begin{figure}[hbtp]
\includegraphics[width=0.4\textwidth]{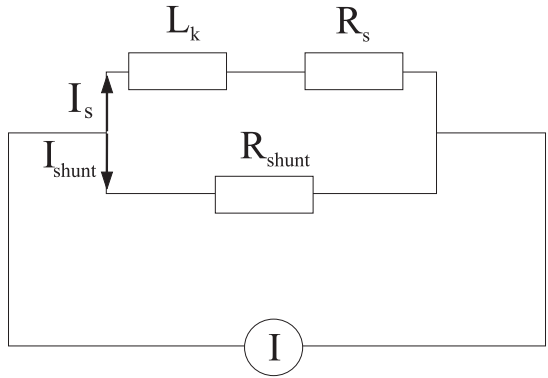}
\caption{The equivalent scheme of the superconducting detector.
The superconductor is modelled by kinetic inductance $L_k$ and
resistance $R_s$ which appeared due to absorbing the photon. The
shunt has resistance $R_{shunt}$.}
\end{figure}

To model real experiments we consider electrical scheme which is
shown in Fig. 2. Here $L_k$ is the kinetic inductance of the
superconducting film, $R_s$ corresponds to the resistance of the
superconductor in the resistive state (in the model geometry - see
Fig. 1) and $R_{shunt}$ is the shunting resistance. For this case
we have to find current $I_s$ which flows via superconductor from
the solution of the following equation
\begin{equation}
\frac{L_k}{c^2}\frac{d I_s}{d t}=(I-I_s)R_{shunt}-V_s
\end{equation}
where the voltage drop over superconductor $V_s$ (over the blue
region in Fig. 1) should be found from the solution of Eq. (4)
with boundary condition $\partial \varphi/\partial x |_{x=\pm
L/2}=-\rho_n I_s/(wd)$.

In numerical calculations we use the dimensionless units. The
order parameter is scaled in units of $\Delta_{GL}(0)$,
temperature is in units of $T_c$ and coordinate is in units of
$\xi_{GL}(0)$. Time is scaled in units of
$\tau_0=\pi\hbar/8k_BT_cu$, electrostatic potential is in units of
$\varphi_0=\hbar/2e \tau_0$ and current density is in units of
$j_0=\hbar/2e \rho_n \tau_0\xi_{GL}(0)$ (depairing current density
in these units is $j_{dep}/j_0=(4/27)^{1/2}(1-T/T_c)^{3/2}$).
\begin{figure}[hbtp]
\includegraphics[width=0.4\textwidth]{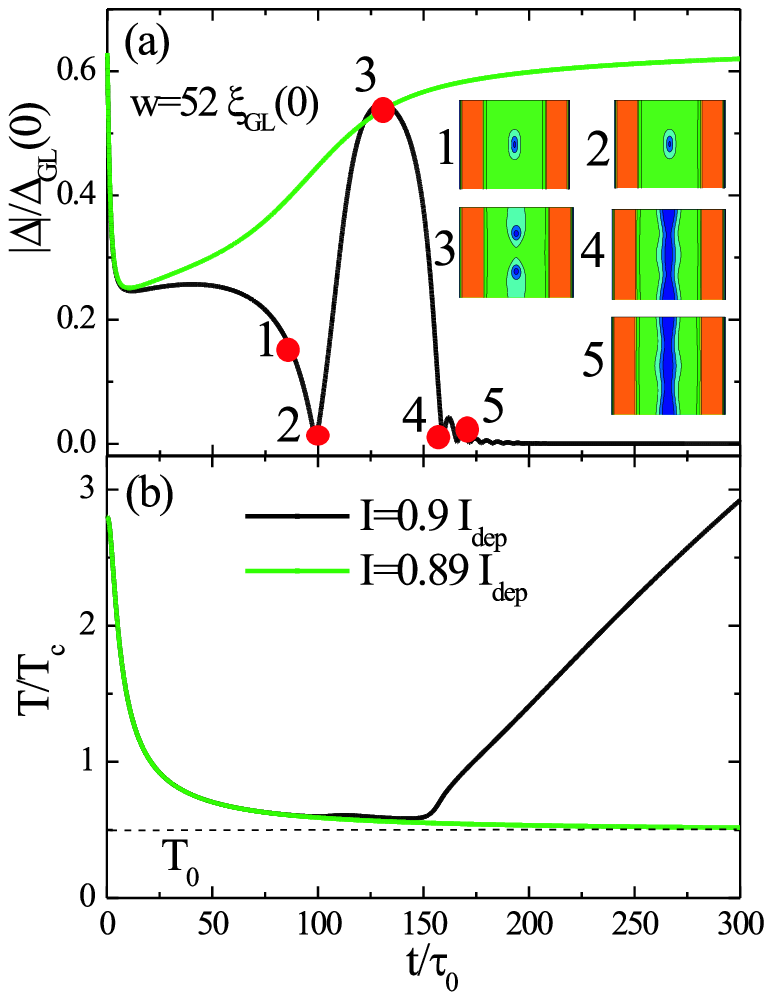}
\caption{Time dependence of the magnitude of the order parameter
(a) and temperature (b) in the center of the hot spot (which
coincide with a center of the film) for two values of the
transport current $I=0.89 I_{dep}$ and $I=0.9 I_{dep}$. The width
of the film $w=52\xi_{GL}(0)$, the local initial increase of the
temperature $\Delta T=2.3T_c$ ($\lambda \simeq 6.5 \mu m$). In the
inset we show contour plots of the magnitude of the order
parameter in the film at different times marked by the numbers on
the black solid curve.}
\end{figure}

To solve Eqs. (2,3,5) numerically we use Euler method and to solve
Eq. (4) - Fourier analysis and cyclic reduction method. In
numerical calculations we first apply the finite current and wait
until all relaxation processes connected with the current induced
suppression of the order parameter stops. Than at some moment of
time we instantly increase the temperature by $\Delta T$ in the
circle or semicircle area inside the superconductor (see Fig. 1)
and studied the dynamical response of the system. The parameters
of the film are: length $l=60 \xi_{GL}(0)$ and width w is varied
from 13$\xi_{GL}(0)$ up to 78 $\xi_{GL}(0)$.

In our calculations we use parameters typical for NbN SSPD
\cite{Semenov1}: $C_v=2.4 mJ cm^{-3} K^{-1}$ , $\tau_{e-e} = 7
ps$, $D=0.45 cm^2$/s, $\xi_{GL}(0)=7.5 nm$, $T_c=10 K$,
$\tau_{e-ph}=17$ ps. At these parameters $L_{e-e}\simeq 18 nm$ and
$\tau_0= 0.052 ps$. In test calculations we consider two values
for $R_{init} = 18 nm$ and $R_{init} = 9 nm$ which are close to
$L_{e-e}$ and found that the results (in particular the value of
the threshold when the voltage appears) differ only slightly. The
presented below results are obtained with $R_{init}=9 nm$. For
this radius and thickness of the film $d=4 nm$ the range of
$\Delta T= 0.3 - 12.8 T_c$ corresponds to the wavelengths $\lambda
\simeq 1.3 - 50 \mu m$. The bath temperature $T_0$ is equal to
$T_c$/2.

\section{Results}

\subsection{Regime with constant current}

At first we consider current bias regime (when $I_s=I$ and
$I_{shunt}=0$ in Fig. 2 for $R_{shunt} \to \infty$). In Fig. 3 we
present time dependence of the magnitude of the order parameter
and effective temperature of quasiparticles in the center of the
film with width $w=52 \xi_{GL}(0)$ and for the situation depicted
in Fig. 1(a) for two close values of the transport current. Notice
that suppression of the order parameter in the center of the hot
spot needs finite time (see Fig. 3(a)). During this time the local
temperature in the center of hot spot decreases (see Fig. 3(b))
due to diffusion of the nonequilibrium quasiparticles and energy
transfer to phonons. When the current is smaller than the
threshold value (we call it the detecting current $I_d$) the order
parameter after reaching some minimal value starts to grow. In
this case the time averaged voltage response is zero. The larger
current destroys superconducting state. In this case $|\Delta|$
oscillates in the center of the hot spot with the amplitude which
decays in time. Each oscillation of $|\Delta|$ corresponds to
nucleation of one vortex-antivortex pair. Motion of the
vortex/antivortex in opposite directions (see inset in Fig. 3(a))
heats the superconductor via Joule dissipation and the local
temperature increases. It results in appearance of the growing
resistive domain (see inset in Fig. 3(a)) in the regime of the
constant current at chosen parameters.

In Figs. 4-5 we show the time evolution of the order parameter in
the films with smaller width at $I>I_d$ and one can see
qualitatively the same scenario of the order parameter dynamics.
The dependence of the detecting current on $\Delta T$ (i.e. on the
energy of the absorbed photon) and width of the film is shown in
Fig. 6. The detecting current decreases with the increase of the
photon energy and its value depends on the position where the
photon is absorbed (in the center or on the edge of the film). For
fixed photon energy the ratio $I_d/I_{dep}$ initially grows with
increasing width of the film and than saturates for large w (see
Fig. 6(b)).

\begin{figure}[hbtp]
\includegraphics[width=0.4\textwidth]{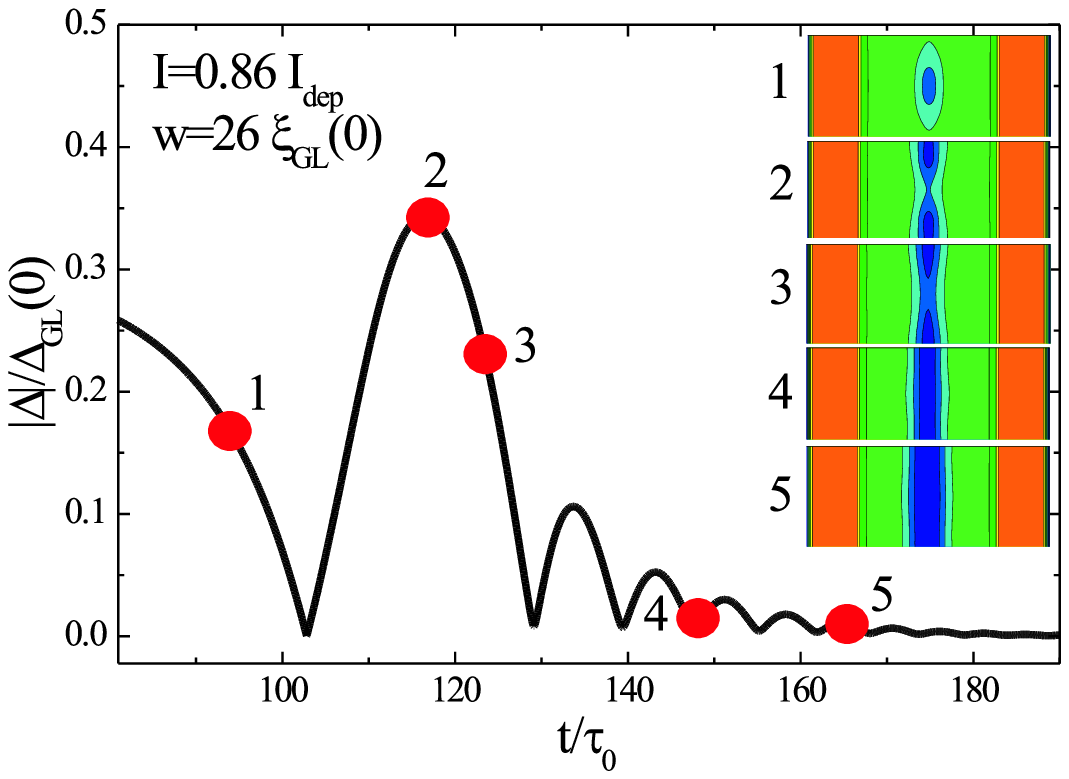}
\caption{Time dependence of the magnitude of the order parameter
in the center of the hot spot (which coincides with the center of
the film). The width of the film $w=26\xi_{GL}(0)$, the bias
current I=0.86 $I_{dep}$, the local initial increase of the
temperature $\Delta T=2.3T_c$ ($\lambda \simeq 6.5 \mu m$). The
inset show contour plots of the magnitude of the order parameter
in the film at different times marked by the numbers on the black
solid curve.}
\end{figure}

\begin{figure}[hbtp]
\includegraphics[width=0.4\textwidth]{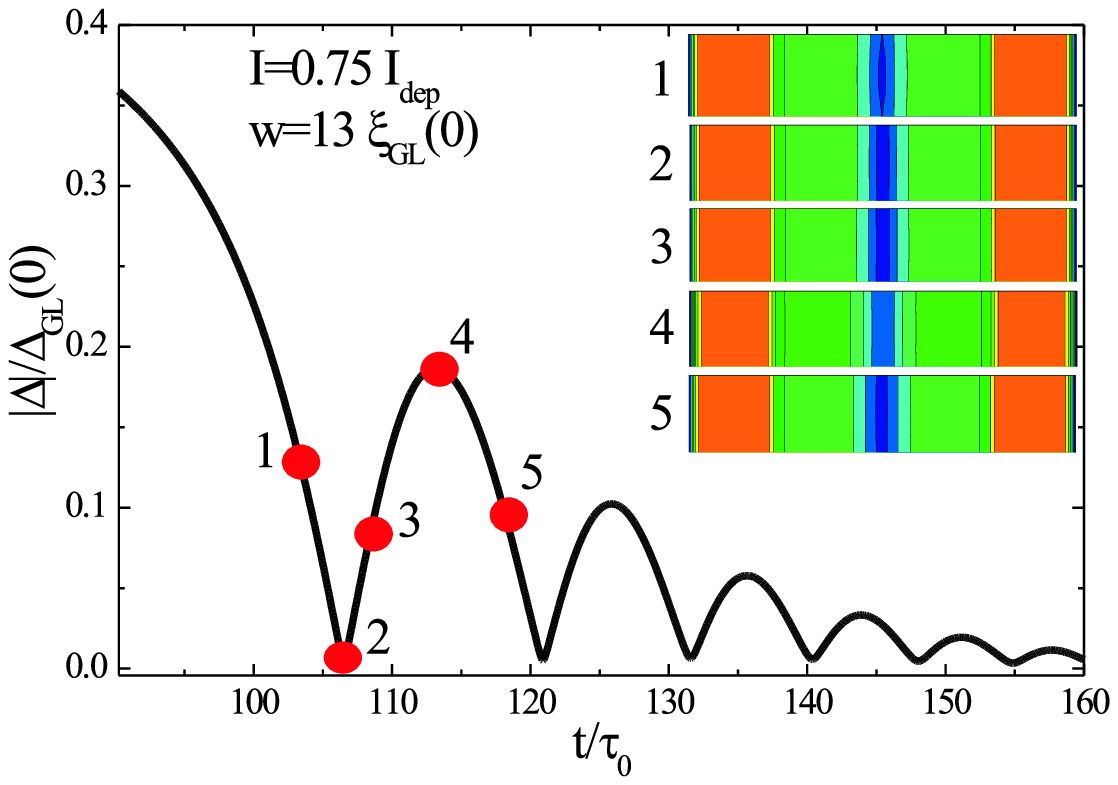}
\caption{Time dependence of the magnitude of the order parameter
in the center of the hot spot (which coincides with the center of
the film). The width of the film $w=13\xi_{GL}(0)$, the bias
current $I=0.75I_{dep}$, the local initial increase of the
temperature $\Delta T=2.3T_c$ ($\lambda \simeq 6.5 \mu m$). The
inset show contour plots of the magnitude of the order parameter
in the film at different times marked by the numbers on the black
solid curve.}
\end{figure}

\begin{figure}[hbtp]
\includegraphics[width=0.45\textwidth]{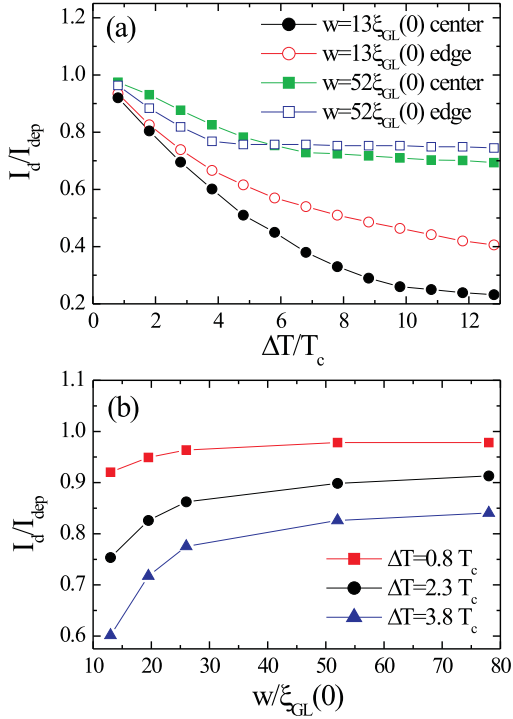}
\caption{(a) The dependence of the detecting current on the
instant increase of the temperature in the circle with radius
$R_{init}$ for narrow and wide films with two positions of the
photon absorbtion (in the center and on the edge of the film). (b)
The dependence of the detecting current on the width of the
superconductive film for three values of $\Delta T/T_c$=0.8, 2.3,
3.8, corresponding to three wavelengths of the electromagnetic
radiation $\lambda=$18.8, 6.5 and 3.9 $\mu m$ respectively (photon
is absorbed in the center of the film).}
\end{figure}

We shall note that for high energy photons (large $\Delta T$) and
relatively narrow film the detecting current is much smaller than
depairing current (see Fig. 6(a)). In this case the Joule
dissipation could be weak, the normal domain does not appear and
superconductivity recovers after nucleation of several
vortex-antivortex pairs in the hot spot area. For film with $w=13
\xi_{GL}(0)$ and $\Delta T= 12.8 T_c$ the normal domain appears
only at $I>0.36 I_{dep}$ ($I_d\simeq 0.23 I_{dep}$ - see Fig.
6(a)) which is close to the value of the current when the heat
dissipation and heat removal are equal to each other
\begin{equation}
\frac{\rho_n j_{heat}^2}{C_v}=\frac{T_c-T_0}{\tau_{e-ph}}
\end{equation}
and for our choice of parameters $I_{heat}=j_{heat}wd\simeq 0.23
I_{dep}$.

According to our numerical calculations the voltage response
appears when the vortex-antivortex pair is nucleated in the center
of the hot spot. To get insight why it occurs we consider the
following simple model. Let us model the region with suppressed
$|\Delta|$ in the hot spot by circle of radius $R$ and we assume
that $|\Delta|$ is spatially uniform and has value $\Delta_{in}$
inside the circle and in the rest of the infinite thin
superconducting film $|\Delta|=\Delta_{out}>\Delta_{in}$. We are
interested how transport current is distributed in such a
superconducting system and when superconducting {\it vortex free}
state becomes unstable. For simplicity we neglect the proximity
effect (which is reasonable when $R\gg \xi$) and use the London
model $j_s=|\Delta|^2\nabla \phi /(4ek_BT_c\rho_n)$. Distribution
of the superconducting current can be found from the current
conservation $div j_s=0$. As a result we obtain that inside the
circle the supervelocity $v\sim \nabla \phi$ is larger than $v$ in
infinity
\begin{equation}
v_{in}=\frac{2v_\infty}{1+\gamma^2}
\end{equation}
($\gamma=\Delta_{in}/\Delta_{out}$) and it is locally enhanced
outside the circle
\begin{equation}
v_{out}(r)=v_\infty\left(1+
\frac{R^2}{r^2}\frac{1-\gamma^2}{1+\gamma^2} \right), r>R
\end{equation}
where the distance is measured from the center of the circle and
we present result at angle $\alpha=\pi/2$ between the direction of
the current and radial vector in polar system of coordinate.

One may find corrections to Eqs. (7,8) for film with finite width
in the limit when $2R/w \ll 1$ and the circle is placed in the
center of the film (see Fig. 1(a)). Assume that for finite film
with $2R/W \ll 1$ the Eqs. (7,8) are approximately valid, but
coefficient $v_{\infty}$ we replace by unknown $v^*$ which we find
from the conservation of the full current
\begin{eqnarray}
I\sim\Delta_{out}^2v_{\infty}wd=2d\int_{0}^{R}\frac{2\Delta_{in}^2v^*}{1+\gamma^2}
dy+ \\
\nonumber +2d\int_{R}^{w/2}\Delta_{out}^2v^*
\left(1+\frac{R^2}{y^2}\frac{1-\gamma^2}{1+\gamma^2} \right) dy
\end{eqnarray}

As a result we find
\begin{equation}
v^*=v_{\infty}/\left(1-\left(\frac{2R}{w}\right)^2\frac{1-\gamma^2}{1+\gamma^2}\right)
\end{equation}
Note that Eq. (10) is also valid (with replacement $2R/w \to R/w$)
for the case when semicircle of radius R with suppressed
$|\Delta|=\Delta_{in}$ is placed on the edge of the film (see Fig.
1(b)). We have to stress that coefficient in front of term
$(2R/w)^2$ in Eq. (10) is approximately valid (up to coefficient
of order of unity) and correct value should be found from the
expansion of the exact result in series with small parameter
$2R/w$.

Because $\Delta_{in}<\Delta_{out}$ and $v_{in}>v_{out}$ the
superconducting Meissner (vortex free) state {\it first} becomes
unstable inside the circle(semicircle). Using value of the
critical supervelocity $v_c\sim |\Delta|$ for instability of
spatially uniform superconducting state which follows from
stationary Eq. (3) we find (by equalizing $v_{in}=v_c$ and using
Eq. (10))
\begin{equation}
\frac{I_{pair}}{I_{dep}}=\frac{\gamma(1+\gamma^2)}{2}\left(1-\left(\frac{2R}{w}\right)^2
\frac{1-\gamma^2}{1+\gamma^2}\right)
\end{equation}
The current $I_{pair}$ is our estimation for the current when the
vortex-antivortex pair is nucleated inside the circle (with
replacement $2R/w\to R/w$ it corresponds to the threshold current
when single vortex is nucleated in the center of the semicircle on
the edge of the film). It is worth to mention here that it is not
the current when the resistive state appears in the sample,
because to escape the circle(semicircle) the vortex and antivortex
should overcome the energy barrier connected with jump in
$|\Delta|$. To estimate this critical current we assume that the
vortices may leave the circle(semicircle) when the supervelocity
(averaged over finite region $\sim \xi(T)$ near the edge of the
circle) is equal to $v_c$. Using Eqs. (8, 10) it is easy to find
that
\begin{equation}
\frac{I_{res}}{I_{dep}}=\left(1-\left(\frac{2R}{w}\right)^2
\frac{1-\gamma^2}{1+\gamma^2}\right)/\left(1+\frac{R}{R+\xi(T)}\frac{1-\gamma^2}{1+\gamma^2}\right)
\end{equation}

One can see that $I_{pair}\leq I_{res}$ (they are equal when
$\Delta_{in}=\Delta_{out}$ and $\gamma=1$). Both critical currents
decrease with decreasing $\Delta_{in}$ and $I_{pair}=0$ and
$I_{res}=I_{dep}(1-4R^2/w^2)/2$ when $R\gg \xi(T)$ and $\gamma=0$
(it corresponds to the normal state of the circle).

To complete the analytical analysis we have to correlate the
radius of the region with suppressed $|\Delta|$ and $\Delta_{in}$
with energy of the incoming photon. Let us assume that the spatial
and time dependence of the temperature after photon absorbtion is
described by the following expression
\begin{equation}
T(r,t)=\frac{\beta}{4\pi Dt}e^{-r^2/4Dt}+T_0
\end{equation}
which is solution of the Eq. (2) with replacement of heating term
by $\beta \delta(t)\delta({\overrightarrow{r}})$ ($\beta=2\pi
\hbar c/\lambda C_vd$) which describes the energy delivered by
photon to the quasiparticles  at the moment t=0 and in the point
r=0 (we also neglect the last term in Eq. (2) because we are
interested in time  interval of about $\tau_{|\Delta|} \ll
\tau_{e-ph}$ after photon absorption).

Local enhancement of temperature leads to suppression of the order
parameter in the hot spot. We may estimate it by using Eq. (3)
where we neglect for simplicity the term with the second
derivative
\begin{equation}
\tau_{|\Delta|}(0)\frac{\partial |\Delta|}{\partial
t}=\left(1-\frac{T}{T_c}-\frac{|\Delta|^2}{\Delta_{GL}(0)^2}\right)
|\Delta|
\end{equation}

One can see that while left hand side of Eq. (14) is negative the
order parameter decreases. Because $T$ is maximal in the center of
hot spot and decreases in time it is reasonable to suppose that
the order parameter stops to decrease when $T=T_c$ in the center
of the hot spot. Using Eq. (13) we find that it occurs at
\begin{equation}
\delta t=\frac{\beta}{4\pi D(T_c-T_0)}\simeq\frac{\Delta
T}{T_c}\frac{\tau_{|\Delta|}(T_0)}{4}
\end{equation}
where we used Eq. (1) to express $\beta=\Delta T \pi R_{init}^2$
and $R_{init}=1.2\xi_{GL}(0)$ via parameters of our numerical
model. Using this result and Eq. (13) we may estimate the size of
the region where the order parameter is suppressed
\begin{equation}
R\simeq 2\sqrt{D\delta t}=\sqrt{\frac{\beta}{\pi(T_c-T_0)}}\simeq
\xi(T)\sqrt{\frac{\Delta T}{T_c}}
\end{equation}
From Eq. (13) one may find suppression of $|\Delta|$ in hot spot
at $r<R$ by moment $t=\delta t$. Below we assume that
$\Delta_{in}=|\Delta|(r=\xi,t=\delta t)$. Using Eqs. (13,14) we
find
\begin{eqnarray}
\gamma=\frac{\Delta_{in}}{\Delta_{out}}\simeq exp\left(
-\frac{1}{\tau_{|\Delta|}(0)} \int_{0}^{\delta
t}\frac{T(\xi,t)}{T_c}dt\right)
\\
\nonumber \simeq
exp\left(-\frac{\beta}{4\pi\xi_{GL}(0)^2T_c}ln\left(\frac{\beta}{\pi\xi_{GL}(0)^2T_c}\right)\right)
\end{eqnarray}
which is approximately valid for photons with
($\beta/\pi\xi_{GL}(0)^2T_c\simeq \Delta T/T_c \gtrsim 1$).

For photon absorbed on the edge (which creates the semicircle)
above results are also valid with the replacement of $\beta$ by
2$\beta$. Finite width of the film does not affect above results
too much while $w\gg 2R$ (or $w\gg R$ for the semicircle) due to
exponential decay of temperature at $r>R$.

Combination of Eqs. (16,17) and Eqs. (11,12) qualitatively
explains our numerical results. First of all with increase of the
photon energy $R$ increases and $\gamma$ decreases providing
decrease of $I_{res}$ and $I_{pair}$ (compare with Fig. 6(a)).

Secondly, above analytical results also explain decrease of
detecting current with decreasing width of the sample (compare
Fig. 6 and Eq. (12)). Thirdly, when the photon is absorbed on the
edge of the film it creates the semicircle with larger radius
$R'=\sqrt{2}R$ (in comparison with photon absorbed in the center).
It results in larger detecting current than $I_d$ for photon
absorbed in the center of the film (see Eqs. (11,12) with
replacement $2R/w$ by $R'/w$ and for high energy photons when
$\gamma \ll 1$). Note that the effect is stronger for films with
smaller width (compare with Fig. 6(a)).

For photons of relatively small energy ($\Delta T/T_c \gtrsim 1$)
which create the circle(semicircle) with small effective radius
$R\simeq \xi(T)$ and wide film ($w\gg \xi(T)$) the situation is
different. In this limit the correction factor due to finite $w$
in Eqs. (11,12) is small and one can see that $I_d$ is smaller for
the photon absorbed on the edge than for the photon absorbed in
the center of the film (due to difference in radiuses and in
$\gamma$ which is finite). It correlates with our numerical
results for low energy photons (small $\Delta T$) and wide film
(see Fig. 6(a)).

We have to note that because of temperature gradient and proximity
effect the distribution of the order parameter is nonuniform in
the hot spot formed by the photon. This factor was not taken into
account in above model and it brings the quantitative difference
between our numerical and analytical results. Our numerical
calculations show that at any considered photon energy the order
parameter is finite in the hot spot at the moment when the first
vortex-antivortex pair is nucleated. After nucleation the vortex
and antivortex becomes immediately unbound and can move freely
across the superconducting film. Therefore found above values for
$I_{pair}$ and $I_{res}$ could be considered as low and upper
thresholds for true detecting current.

\subsection{Regime with changing current}

In the experiments the circuity shown in Fig. 2 is usually used to
prevent the latching of the superconductor in the normal state. As
a result with appearance of the voltage drop over superconductor
the current via the superconducting sample decreases and it
switches back to the superconducting state.

In our calculations we use $R_{shunt}$=50 Ohm. The kinetic
inductance is evaluated using the following expression:
\begin{equation}
L_k=\frac{4 \pi \lambda_L^2 l}{wd}
\end{equation}
where $\lambda_L$ is a London penetration depth.
\begin{figure}[hbtp]
\includegraphics[width=0.4\textwidth]{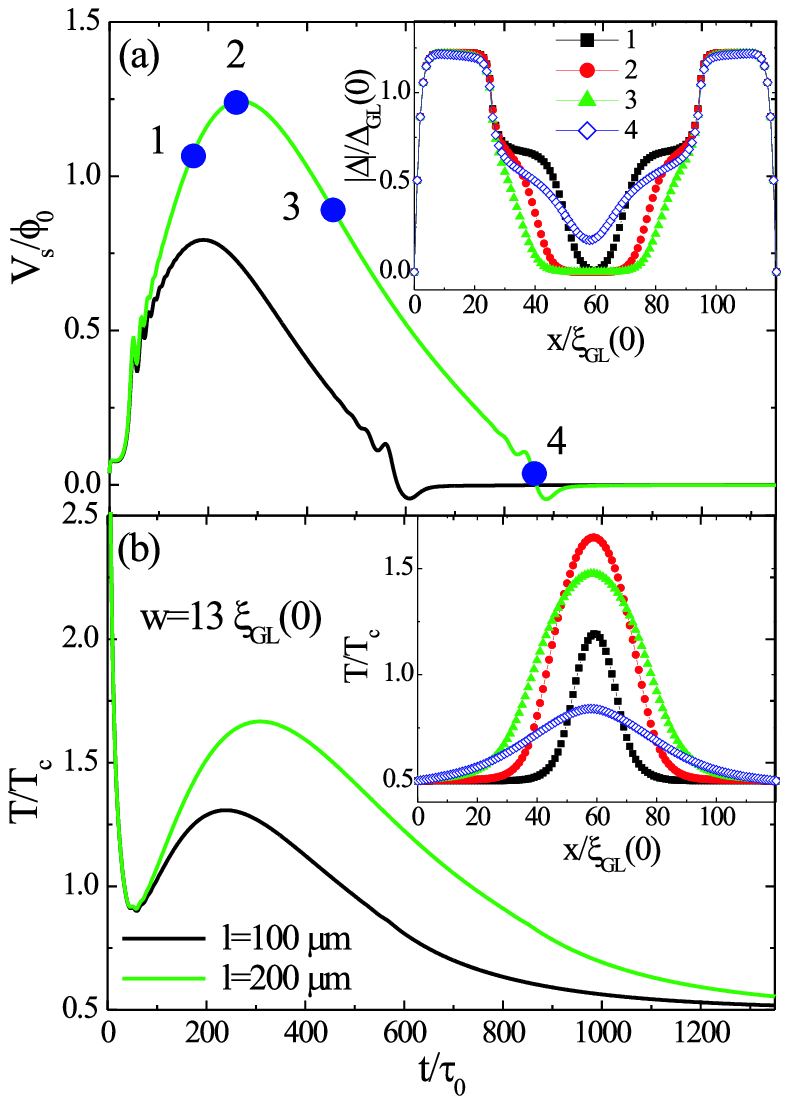}
\caption{(a) The time dependence of the voltage drop via the
superconductive film calculated for two lengths: l=100 $\mu m$ and
l=200 $\mu m$. The inset shows the distribution of the magnitude
of the order parameter along the film (at y=0) at different
moments in time. (b) The time dependence of the temperature in the
center of the film (hot spot) for the same films. The inset shows
the distribution of the temperature along the film (at y=0) at
different moments in time. The width of the film
$w=13\xi_{GL}(0)$, the bias current $I=0.6I_{dep}$, the local
initial increase of the temperature $\Delta T=3.8T_c$ ($\lambda
\simeq 3.9 \mu m$).}
\end{figure}

In Fig. 7(a) we present the time dependence of the voltage drop
$V_s$ via the superconductor (resistance $R_s$ in Fig. 2) for two
films with lengths l=100 $\mu m$ and 200 $\mu m$ ($w$=100 nm,
$d$=4 nm and $\lambda_L=400$ nm \cite{Kamplapure}). It is seen
that with decrease of the kinetic inductance the duration and the
amplitude of the voltage pulse becomes shorter and smaller
correspondingly. The reason is simple - for smaller $L_k$ current
via superconductor decreases faster, the temperature inside the
normal domain increases slower (see Fig. 7(b)) and it takes less
time to cool the sample up to bath temperature $T_0$ (see Fig.
7(b)).

For the longest film the duration of the voltage pulse is about
900 $\tau_0$ which is much larger than the typical time interval
between consequent nucleation of the vortex-antivortex pair
$\Delta t \simeq 10 \tau_0$ (we roughly estimate it as a time
interval between nucleation of the second and third
vortex-antivortex pairs - see Fig. 5). Therefore at least 90
vortex-antivortex pairs are nucleated during this voltage pulse.

\begin{figure}[hbtp]
\includegraphics[width=0.4\textwidth]{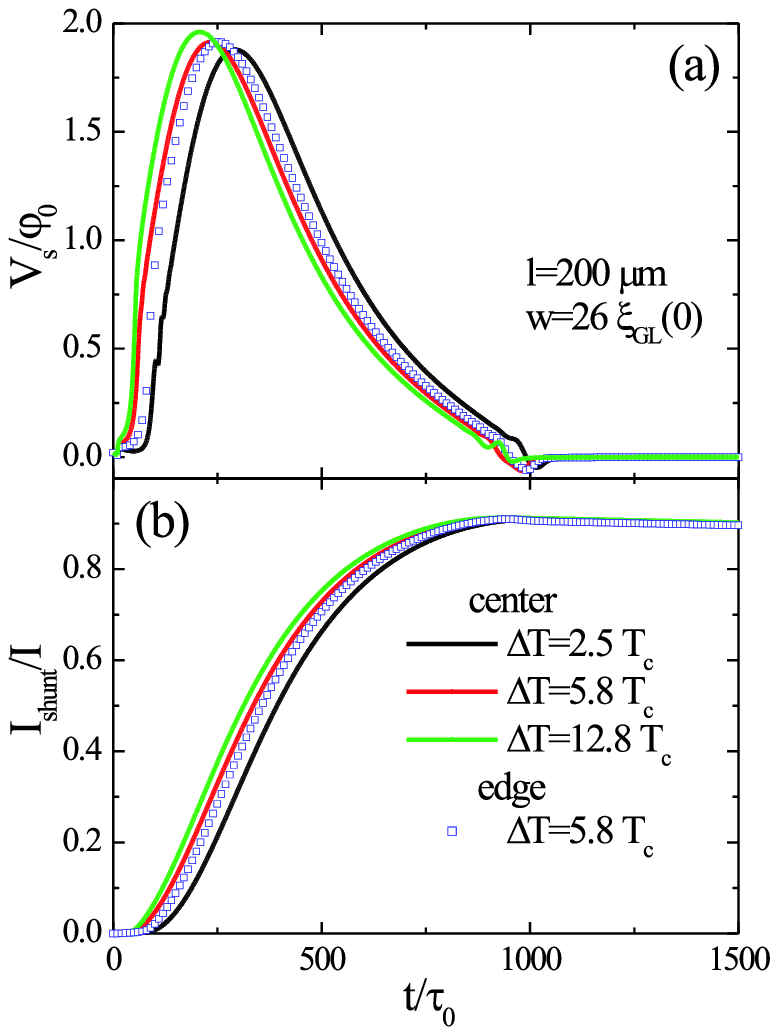}
\caption{(a) The time dependence of the voltage via the
superconductive film with length l=200 $\mu m$ and width $w=26
\xi_{GL}(0)$ for photons of different energy (different $\Delta
T$) and position of the absorbtion (in the center or on the edge
of the film). (b) The time dependence of the current via shunt
resistance (the bias current $I=0.88I_{dep}$).}
\end{figure}

In Fig. 8(a) we present the time dependence of $V_s$ for photons
of different energy and position where it is absorbed. Notice that
the shape of the voltage pulse and current $I_{shunt}$ (see Fig.
8(b)) via shunt resistance (which is measured in the experiments
with SSPD) slightly depend both on $\Delta T$ and absorption
position. At times larger than $1000 \tau_0$ when the voltage drop
via superconductor is equal to zero (see Fig. 8(a)) the current
$I_{shunt}$ decays with characteristic time
$\tau=L_k/R_{shunt}c^2\simeq 5 \cdot 10^4 \tau_0$ for our choice
of parameters.

We have to stress that our results for voltage pulse via
superconductor (duration of the pulse, its amplitude) are obtained
with neglecting the back action of the 'heated' phonons on the
electrons. For our parameters ($\tau_{e-ph}\simeq 327 \tau_0$) the
duration of the pulse is about $3\tau_{e-ph}$. But heating of the
phonon subsystem occurs over time $t\gg \tau_{e-ph}$
\cite{Semenov3}. Therefore we expect that hot phonons do not
strongly influence the duration of the voltage pulse and its
amplitude. And it certainly should not influence the value of the
detecting current because it is determined from the stability of
the superconducting state with respect to appearance of the single
vortex or vortex-antivortex pair during evolution of the photon
induced hot spot.

\section{Relation to an experiment}

The model which we use (Eqs. (1-4)) is strongly oversimplified.
First of all it does not take into account lose of the energy of
photon at initial stage of formation of the hot spot on time scale
$\tau_{e-e}$. Secondly we neglect temperature dependence of $C_v$,
oversimplify the energy transfer to the phonons (last term in Eq.
(2)) and we did not take into account possibility of direct
partial destruction of the superconducting order parameter by
incoming photon (in our approach $|\Delta|$ is influenced only via
effective temperature of quasiparticles). Moreover, the
time-dependent Ginzburg-Landau equation (Eq. (3)) is not
quantitatively correct at temperatures lower than $\sim 0.9 T_c$
and (within quasiequilibrium approach) when $\tau_{e-e} \gtrsim
\tau_{|\Delta|}$. Therefore direct {\it quantitative} comparison
of our results with experiment (at least at low temperatures)
looks speculative.

But despite this we believe that the used model catches the main
physical mechanism of photon detection by current-carrying
superconducting film which is the following. The incoming photon
partially suppresses order parameter in the finite region. It
leads to redistribution of the current density (supervelocity) in
the film and such a state becomes unstable ({\it without any
fluctuations and if current is large enough but smaller than
depairing current}) with respect of appearance of the unbound
vortex-antivortex pair (if photon is absorbed far from edges) or
single vortex (if photon is absorbed near the edge of the film).
Lorentz force causes motion of these vortices that heats the film
locally and gives rise to a voltage pulse.

We hope that our results could be used for understanding (not for
direct fitting) of some experimental results. For example found
dependence of the detection current on the position of the hot
spot (see Fig. 6(a) and our analytical model) may be used for
qualitative explanation of the monotonous decrease of the
detection efficiency (DE) with the decrease of the energy of the
incoming photon \cite{Semenov4,Hofherr,Maingault}. Indeed, if we
fix the current (for example on the level $I=I_{dep}/2$ - see Fig.
6(a)) and start to decrease the energy of the photon, the edge
region of the superconducting film first is "switched off" from
the detection process (when $\Delta T \simeq 8T_c$ which
corresponds to $\lambda\simeq 1.9 \mu m$ - see Fig. 6(a)) and at
$\Delta T \simeq 5 T_c$ ($\lambda\simeq 2.9 \mu m$) the central
region of the superconducting film stops to detect photons. As a
result there is finite range of the wavelengths $\Delta \lambda
\simeq 1 \mu m$ at ($I=I_{dep}/2$) where the detection efficiency
gradually changes (qualitatively such a behavior was observed in
Refs. \cite{Semenov4,Hofherr,Maingault}).

If we fix the energy of photon then with increase of the current
the central region of the film first starts to detect the photons
and than the edge regions join the detection process. Therefore
there is a finite interval of the currents ($\delta I$) within
which DE gradually grows with current increase (qualitatively such
a behavior was observed in Ref. \cite{Baek}). The real samples
have different kinds of imperfections (variations of the
thickness, width, material parameters, turns) having their own
values of the detecting current and it obviously affects $\delta
I$ at low currents. Our result show that even in ideal samples
with no imperfections and at low temperatures (when effect of
fluctuations is rather small) there will be finite $\delta I$
connected with presence of the edges. And our prediction is that
the wider the sample the narrower is this interval of currents
(see Fig. 6(a) and Eq. (12)).

\section{Conclusion}

In our work we use the quasiequilibrium approach and describe the
deviation from the equilibrium in terms of the effective
temperature of the quasiparticles which depends on time and
coordinate. We assume that the absorbed photon creates initially
the hot spot in the superconducting film with radius $R_{init}$
and local enhancement of the quasipartcile temperature by $\Delta
T$ which is proportional to the energy of the photon. The temporal
and spatial evolution of the effective temperature and
superconducting order parameter in the superconductor we study by
numerical solution of the time-dependent Ginzburg-Landau equation
coupled with Poisson's equation for an electrical potential and
heat diffusion equation.

We show that for photon of fixed energy the voltage response
appears only at the current larger some critical one (we call it
as detecting current $I_d$) when the vortex-antivortex pair is
nucleated in the center of the photon induced hot spot. Motion of
the vortex and antivortex in opposite directions heats the
superconductor and leads to the appearance of the growing normal
domain when detecting current $I_d$ is about of depairing current.
For high energy photon and narrow film $I_d \ll I_{dep}$ the
motion of the vortices-antivortices does not heat the
superconductor and sample goes back to the superconducting state
after nucleation of several vortex-antivortex pairs at $I\simeq
I_d$ even in regime of constant current.

We find numerically that with increasing the width of the
superconducting film $I_d$ increases and stays less than $I_{dep}$
even for infinite film. We also find that the detecting current
for photon absorbed on the edge of the film differs from $I_d$ for
photon absorbed in the center of the film.

We develop simple analytical model to explain above results. We
assume that absorbed photon creates in the superconducting film of
finite width the region (in the form of circle or semicircle)
where the superconducting order parameter $|\Delta|$ is partially
suppressed and in framework of the London model we study the
current redistribution and stability of the current-carrying state
in such a system. We find that the superconducting vortex free
state becomes unstable in the region with suppressed
superconductivity at $I_{pair}<I_{dep}$ and resistive state
appears at larger current $I_{pair}<I_{res}<I_{dep}$. Our
analytical model predicts different values for the detecting
current of the photon absorbed on the edge and in the center of
the film in agreement with our numerical results. The last effect
is connected with different redistribution of the supervelocity
(current density) inside and outside the circle placed in the
center of the film and the semicircle placed on the edge of the
same film. We expect that our analytical results within the London
model are valid for arbitrary temperatures contrary to the results
based on the time-dependent Ginzburg-Landau and heat conductance
equations which are strictly valid at $T>0.9T_c$ and when
$\tau_{e-e}\ll \tau_{|\Delta|}$ and $\tau_{e-e}\ll \tau_{e-ph}$.

To model operation of real superconducting single photon detector
we consider the scheme with the resistance which is switched on in
parallel to the detector and take into account the large kinetic
inductance of real SSPD. We find that duration and amplitude of
voltage pulse decrease with the decrease of the kinetic inductance
while the detecting current practically does not change. We also
find that the shape of the voltage pulse weakly depends on the
energy of the absorbed photon and on the place of the absorbtion
for the homogeneous film.

\begin{acknowledgments}

We acknowledge stimulating discussions with Alexei D. Semenov.
This work was supported by the Russian Foundation for Basic
Research and Russian Agency of Education under the Federal Target
Programme "Scientific and educational personnel of innovative
Russia in 2009-2013".

\end{acknowledgments}

\end{document}